\documentclass[runningheads]{llncs}
\usepackage{graphicx}
\usepackage{graphicx}
\usepackage{amsmath}
\usepackage{float}
\usepackage{hyperref}
\usepackage{algorithm}
\usepackage{algorithm, float}
\usepackage[noend]{algpseudocode}
\usepackage{multirow}
\usepackage{graphicx} 
\usepackage{caption,subcaption}
\usepackage{xcolor}
\usepackage{xcolor}

\newcommand{\xb}{\textbf{x}} 
\newcommand{\yb}{\textbf{y}} 
\newcommand{\Eb}{\textbf{E}} 

\newcommand{\Bb}{\textbf{B}} 

\makeatletter
\newcommand{\printfnsymbol}[1]{%
  \textsuperscript{\@fnsymbol{#1}}%
}
\makeatother

\begin{document}
\title{Joint reconstruction and bias field correction for undersampled MR imaging\thanks{This work was partially funded by grant 205321\_173016 Schweizerischer Nationalfonds zur Förderung der Wissenschaftlichen Forschung}}

%
\author{Mélanie Gaillochet\thanks{Equal contribution.} \and
Kerem C. Tezcan\printfnsymbol{2} \and
Ender Konukoglu}
%
\authorrunning{M. Gaillochet et al.}
%
\institute{Computer Vision Lab, ETH Zürich, Zürich, Switzerland\\ 
\email{gamelani@student.ethz.ch}\\
\email{\{tezcan, kender\}@vision.ee.ethz.ch}}


\maketitle              
\begin{abstract}
Undersampling the k-space in MRI allows saving precious acquisition time, yet results in an ill-posed inversion problem. Recently,  many deep learning techniques have been developed, addressing this issue of recovering the fully sampled MR image from the undersampled data. However, these learning based schemes are susceptible to differences between the training data and the image to be reconstructed at test time. One such difference can be attributed to the bias field present in MR images, caused by field inhomogeneities and coil sensitivities. In this work, we address the sensitivity of the reconstruction problem to the bias field and propose to model it explicitly in the reconstruction, in order to decrease this sensitivity. To this end, we use an unsupervised learning based reconstruction algorithm as our basis and combine it with a N4-based bias field estimation method, in a joint optimization scheme. We use the HCP dataset as well as in-house measured images for the evaluations. We show that the proposed method improves the reconstruction quality, both visually and in terms of RMSE.

\end{abstract}

\section{Introduction}

Magnetic resonance imaging (MRI) is a non-invasive imaging technique that allows studying anatomy and tissue properties without ionizing radiation. However, acquiring a detailed high-quality image is time-consuming. 

Reducing scan time is therefore essential in order to increase patient comfort and throughput, and to open up more possibilities for optimized examinations. Since acquisition time in MRI is directly related to the number of samples collected in k-space, shortening it usually means reducing the number of samples collected - for instance, by skipping phase-encoding lines in k-space in the Cartesian case. Yet doing so violates the Nyquist criterion, causing aliasing artifacts and requiring reconstruction techniques that can recover the fully sampled MR image from the undersampled data.

The problem of designing such techniques has received considerable attention in the clinical and signal processing research communities. Conventional approaches involve compressed sensing \cite{compressed_sensing,cs_MRI} and parallel imaging \cite{parallel_imaging}. More recently, the research efforts have focused on using deep learning to tackle the problem and achieved state-of-the-art results. However, despite their multiple advantages, these learning based algorithms are susceptible to discrepancies between the training set and the image to reconstruct at test time. One such difference is due the different bias fields present in the images. Such discrepancies, sometimes referred to as `domain shift', typically lead to a loss in performance. Various methods have been developed to tackle the domain shift problem in the deep learning community. One approach is to augment the training data to hopefully resemble the test data better~\cite{volpi_generalization}\cite{krishna_ipmi}. Although this approach increases the generalization capabilities of the learned model, it does not optimize the solution specifically for the test image at hand. Another approach is to modify the network parameters to reduce the domain gap~\cite{sun_coral}, which indirectly optimizes the solution by modifying the network rather than the solution.  

Yet, while the effect of the bias field on the image encoding is easy to model and there are readily available methods that can estimate it well, none of the aforementioned approaches takes these into account.

In this work we propose a joint reconstruction algorithm, which estimates and explicitly models the bias field throughout the reconstruction process. By doing so, we remove one degree of variation between the training set and the image to be reconstructed. The training is done on images without bias field, and the bias field itself is modeled as a multiplicative term in the image encoding process, linking the training and test domains. In order to be able to do this we use a reconstruction algorithm which decouples the learned prior information from the image generation process, namely the DDP algorithm~\cite{DDP}. Using the N4 algorithm, we iteratively estimate the bias field in the test image  throughout the reconstruction, and this estimation improves as the reconstructed image become better. We compare the proposed method to reconstruction without bias field estimation, using a publicly available dataset as well as in-house measured images, and show improvement in performance.

\section{Methods}
\subsection{Problem formulation}
The measured k-space data, $\yb\in \mathcal{C}^M$ and the underlying true MR image, $\xb\in \mathcal{C}^N$ are related through the encoding operation $\Eb : \mathcal{C}^N \to \mathcal{C}^M$ (which incorporates the coil sensitivites, Fourier transformation and the undersampling operation) as $\yb = \Eb\xb + \eta$, where $\eta$ is complex Gaussian noise.

\subsection{DDP reconstruction}
The deep density prior reconstruction is based on maximum-a-posteriori estimation, i.e. $\max_\xb \log p(\xb|\yb) = \max_\xb \log p(\yb|\xb) + \log p(\xb)$, where $\log p(y|x)$ is the data consistency and $\log p(x)$ is the prior term. The data consistency can be written exactly as the log Gaussian and the method approximates the prior term using the evidence lower bound (ELBO) of a variational auto-encoder (VAE)~\cite{rezende_vae}\cite{kingma_vae}.
With these, the reconstruction problem becomes
\begin{equation}\label{eqn:vanilla}
    \min_x ||\Eb\xb - \yb || - ELBO(\xb).
\end{equation}
The VAE is trained on patches from fully sampled images and the ELBO term operates on a set of overlapping patches that cover the whole image. As evident in the equation above, the ELBO term is independent of the image encoding.
The DDP method solves the problem using the projection onto convex sets (POCS) algorithm. In this scheme the optimization is implemented as successive applications of projection operations for prior $\mathcal{P}_{prior}$, data consistency $\mathcal{P}_{DC}$ and the phase of the image $\mathcal{P}_{phase}$, i.e. $\xb^{t+1} = \mathcal{P}_{DC}\mathcal{P}_{phase}\mathcal{P}_{prior}\xb^t$. The prior projection is defined as a gradient ascent for a fixed number of steps for the ELBO term w.r.t. the image magnitude, i.e. $\mathcal{P}_{prior}\xb = \xb^N$, where $\xb^{n+1} = \xb^n + \alpha \frac{d}{d\xb} ELBO(|\xb|)|_{\xb=\xb^n}$ for $n=0...N$. The data consistency projection is given as $\mathcal{P}_{DC}\xb = \xb - \Eb^H ( \Eb \xb - \yb)$. For the phase projection, we use the one defined by \cite{DDP}, to which we add the minimization of the data consistency with respect to the image phase. For further details, we refer the reader to the original paper \cite{DDP}.
\subsection{Modeling and estimating the bias field}

Signal intensities are often not constant across the MR images, even inside the same tissue. Instead, they usually vary smoothly, with fluctuations of 10\%-20\%, across the measured 3D volume \cite{BiasField_N3}. These variations, collectively called the bias field, can be attributed to factors like patient anatomy, differences in coil sensitivity or standing wave effects \cite{BiasField_N3,BiasField_MRI_Sled}, which are difficult or impossible to control during an acquisition. Hence they can introduce a degree of variation between images used for training reconstruction models and an image to be reconstructed at test time. In order to prevent a loss of performance, this variation has to be taken into account. 

To this end, we model the bias field explicitly and incorporate it into Equation~\ref{eqn:vanilla} as a multiplicative term $\Bb$ before the encoding operation, modifying the image intensities pixelwise. 
The bias field is an additional unknown in the reconstruction process that is estimated alongside $x$ as

\begin{equation}\label{eqn:withbf}
    \min_{\xb, \Bb} ||\Eb\Bb\xb - \yb || - ELBO(\xb).
\end{equation}
Here, the measured k-space $\yb$ carries the effect of a bias field, while $\xb$ is bias field free as the bias field in the image is explicitly modeled using the $\Bb$ term. This setting allows us to learn the VAE model on images without bias field. The advantage of this idea is two fold. Firstly, since training can be performed on bias field free images, it is easier for the VAE to learn the distribution as there is less spurious variation in the data. 
Secondly, we make the reconstruction problem easier by explicitly providing the bias field information, which otherwise would have to be reconstructed from the undersampled k-space as well. 
We solve Equation~\ref{eqn:withbf} as a joint iterative reconstruction problem by minimizing alternatively two sub-problems:
\begin{align}
    1. \hspace{0.5cm} \xb^t &= \min_{x} ||\Eb\Bb^{t-1}\xb - \yb || - ELBO(\xb) \\
    2. \hspace{0.5cm} \Bb^t &= \text{N4}(\Bb^{t-1}\xb^t),
\end{align}
where N4 denotes the bias field estimation algorithm, which we will explain below. To account for the bias field, the data consistency projection $\mathcal{P}_{DC}$ needs to be adapted and becomes $\mathcal{P}_{DC}^{B}\xb = \Bb^{-1} [ \Bb \xb - \Eb^H ( \Eb \Bb \xb - \yb) ]$. In this case the reconstructed image corresponding to $\yb$ is given as $\Bb\xb$. This modification can be interpreted as doing a forward-backward projection with $\mathcal{P}_{bias} = \Bb$ before and after the data consistency projection to move the image between the ``normalized'' bias field free domain and the bias field corrupted acquisition domain, i.e. $\xb^{t+1} = 
\mathcal{P}_{bias}^{-1}\mathcal{P}_{DC}\mathcal{P}_{bias}\mathcal{P}_{phase}\mathcal{P}_{prior}\xb^t$. The pseudocode for the described joint optimization scheme is presented in Algorithm~\ref{alg:recon}.

\subsubsection{N4 bias field estimation}
N4 is a variant of the widely used N3 algorithm \cite{N4_paper}, a non-parametric iterative method that approximates intensity non-uniformity fields in 3-dimensional images. Given an image, its objective is to find a smooth, slowly varying, multiplicative field~\cite{BiasField_N3}. N4 improves upon the N3 algorithm by modifying the B-spline smoothing strategy and the iterative optimization scheme used in the original framework. We use the implementation available as N4ITK \cite{N4_itk}. We denote this method as $N4(\cdot)$ in our formulations.

\begin{algorithm}[H]
\caption{Joint reconstruction}\label{alg:recon}
\begin{algorithmic}[1]
\State $\yb \gets$ undersampled k-space data
\State \Eb $\gets$ undersampling encoding operator
\State VAE $\gets$ trained VAE
\State NumIter, BiasEstimFreq, DCProjFreq

\Procedure{JointRecon}{y, E, VAE}

\State $\Bb \gets$ N4$(\Eb^H\yb$)
\State $\xb^0 \gets \left(\Bb\right)^{-1}\Eb^H\yb$

\For {$t$:  0 to $\textrm{NumIter - 1}$}
    \State $\xb^{t + 1} = \mathcal{P}_{prior} \xb^t$
    
    \State $\xb^{t + 1} \gets \mathcal{P}_{phase} \xb^{t + 1}$ \Comment{Optional}
    
    
    \If { $t \%$ DCProjFreq $ == 0$ and $t \neq 0$}
        \State $\textrm{\xb}^{t + 1} \gets \Bb^{-1} \left[ \Bb  \xb^{t + 1} - \Eb^H ( \Eb \Bb \xb^{t + 1} - \yb) \right]$ \Comment{$\mathcal{P}_{DC}^B$}
    \EndIf
    
    \If {$t \%$ BiasEstimFreq $ ==0$ and $t \neq 0$}
            \State $\Bb \gets$ N4$(\Bb \xb^{t + 1})$
    \EndIf
\EndFor

\Return $\xb^{t+1}$, $\Bb$

\EndProcedure
\end{algorithmic}
\label{algo_custom}
\end{algorithm}

\subsection{Datasets used}
To train the VAE we used 5 central, non-adjacent slices with 0.7mm isotropic resolution from T1 weighted images of 360 subjects from the Human Connectome Project (HCP) preprocessed dataset \cite{HCP_dataset,HCP_doc}, which by default have a bias field. We used N4 on the images to also create a bias field free training set. 

For test images, we took a central slice from 20 different test subjects from the HCP data. As the HCP images tend to have a similar bias field, we additionally created a modified test set where we estimated the bias fields with N4, took their inverse and multiplied them with the bias field free images. In addition to HCP data, we also tested the proposed method with central slices from 9 in-house measured subjects. These images were acquired using a 16 element head coil and have similar acquisition parameters as the HCP dataset with a 1mm isotropic resolution. We used ESPIRiT~\cite{espirit} to estimate the coil sensitivity maps for these images.

\subsection{Training the VAE}
We trained four patch-wise VAEs - for two different resolution levels to match the datasets, each with and without bias field. We used patches of size 28x28  with a batch size of 50 and ran the training for 500,000 iterations. The patches were extracted randomly from the training images with replacement. All the VAEs were trained with the same training images extracted from the HCP dataset, as described above.

\subsection{Experimental Setup}
We used random Cartesian undersampling patterns with 15 fully sampled central profiles. We generated a different pattern for each subject, and applied the same pattern for a given subject throughout all experiments for comparability of results. When reconstructing test images from the HCP dataset, we used 302 iterations (NumIter) for R=2, 602 for R=3, 1002 for R=4 and 1502 for R=5, to allow for convergence of reconstructed images. Since the in-house measured images have multiple coils,
the successive applications of data consistency projections coincide with a POCS-SENSE reconstruction~\cite{pocssense}, speeding up convergence and requiring less iterations. Hence, when performing reconstruction on images from the in-house measured dataset, we ran the first 10 iterations without prior projection, applying only data consistency projections.
Additionally, the discrepancy between the actual coil sensitivities and the ESPIRiT~\cite{espirit} estimations may lead to divergence after too many iterations. Hence, a lower number of iterations was used for reconstruction experiments on the in-house measure dataset: 32 iters for R=2, 102 for R=3 and R=4, and 202 for R=5.  For all test datasets, the parameters were set as $\alpha$ = 1e-4, BiasEstimFreq=10, DCProjFreq=10. 
As for the N4 bias field estimation algorithm, the default parameters were used.

\begin{figure}[H]
    \centering
    \begin{subfigure}{0.18\textwidth}
        \includegraphics[width=\linewidth]{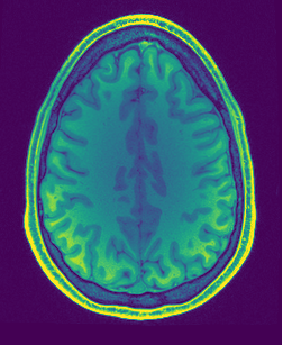}
    \end{subfigure}\hfil 
    \hfil
    \begin{subfigure}{0.18\textwidth}
      \includegraphics[width=\linewidth]{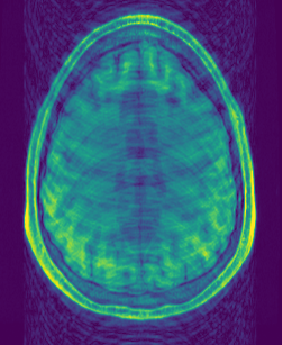}
    \end{subfigure}\hfil 
    \hfil
    \begin{subfigure}{0.18\textwidth}
      \includegraphics[width=\linewidth]{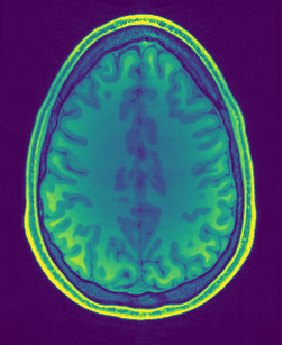}
    \end{subfigure}\hfil 
    \hfil
    \begin{subfigure}{0.18\textwidth}
      \includegraphics[width=\linewidth]{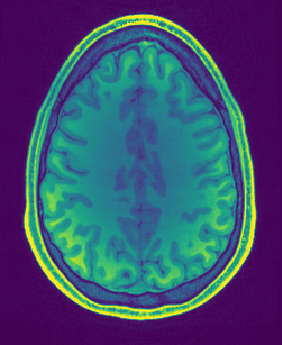}
    \end{subfigure} \hfil 
    \medskip 
    
    \begin{subfigure}{0.18\textwidth}
        \includegraphics[width=\linewidth]{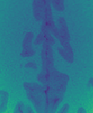}
        
    \end{subfigure}\hfil 
    \hfil
    \begin{subfigure}{0.18\textwidth}
      \includegraphics[width=\linewidth]{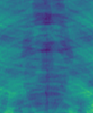}
      
    \end{subfigure}\hfil 
    \hfil
    \begin{subfigure}{0.18\textwidth}
      \includegraphics[width=\linewidth]{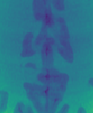}
      
    \end{subfigure} \hfil 
    \hfil
    \begin{subfigure}{0.18\textwidth}
      \includegraphics[width=\linewidth]{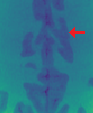}
      
    \end{subfigure} \hfil 
    \medskip
    
     \begin{subfigure}{0.18\textwidth}
        \includegraphics[width=\linewidth]{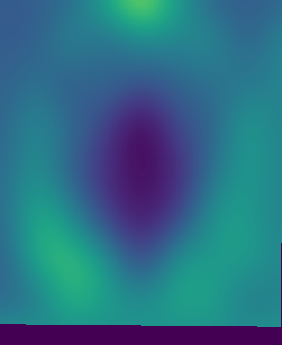}
    \end{subfigure}\hfil 
    \hfil
    \begin{subfigure}{0.18\textwidth}
      \includegraphics[width=\linewidth]{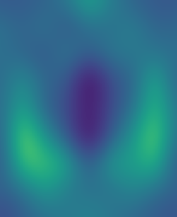}
    \end{subfigure}\hfil 
    \hfil
    \begin{subfigure}{0.18\textwidth}
      \includegraphics[width=\linewidth]{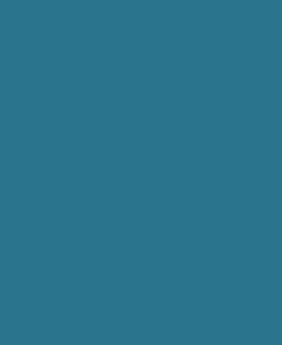}
    \end{subfigure}\hfil 
    \hfil
    \begin{subfigure}{0.18\textwidth}
      \includegraphics[width=\linewidth]{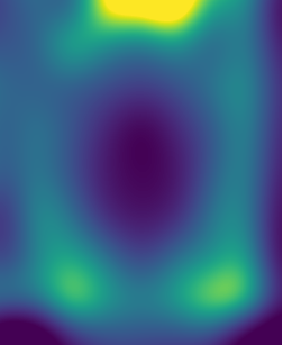}
    \end{subfigure} \hfil 
    \medskip

        \begin{subfigure}{0.18\textwidth}
        \includegraphics[trim={1.5cm 0.8cm 1.0cm 2.1cm},clip,width=\linewidth]{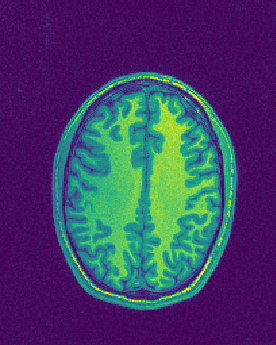}
    \end{subfigure}\hfil 
    \hfil
    \begin{subfigure}{0.18\textwidth}
      \includegraphics[trim={1.5cm 0.8cm 1.0cm 2.1cm},clip,width=\linewidth]{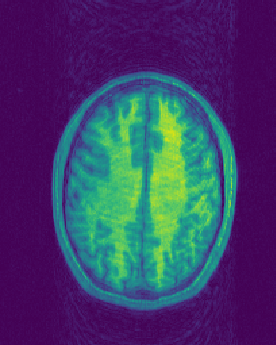}
    \end{subfigure}\hfil 
    \hfil
    \begin{subfigure}{0.18\textwidth}
      \includegraphics[trim={1.5cm 0.8cm 1.0cm 2.1cm},clip,width=\linewidth]{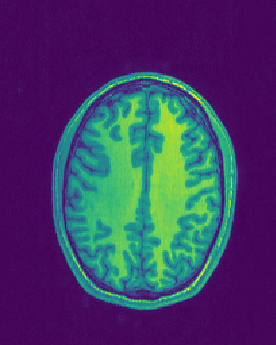}
    \end{subfigure} \hfil 
    \hfil
    \begin{subfigure}{0.18\textwidth}
      \includegraphics[trim={1.5cm 0.8cm 1.0cm 2.1cm},clip,width=\linewidth]{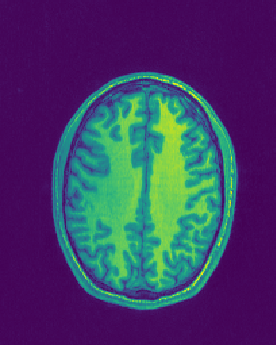}
    \end{subfigure} \hfil 
    \medskip
    
    \begin{subfigure}{0.18\textwidth}
        \includegraphics[trim={1.5cm 0.8cm 1.0cm 2.1cm},clip,width=\linewidth]{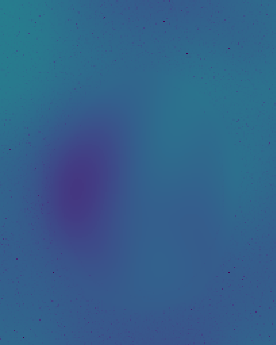}
    \end{subfigure}\hfil 
    \hfil
    \begin{subfigure}{0.18\textwidth}
      \includegraphics[trim={1.5cm 0.8cm 1.0cm 2.1cm},clip,width=\linewidth]{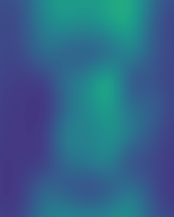}
    \end{subfigure}\hfil 
    \hfil
    \begin{subfigure}{0.18\textwidth}
      \includegraphics[trim={1.5cm 0.8cm 1.0cm 2.1cm},clip,width=\linewidth]{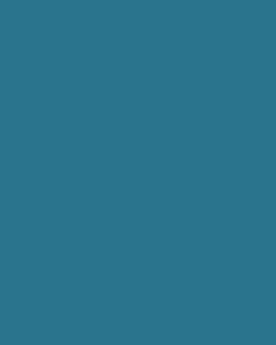}
    \end{subfigure} \hfil 
    \hfil
    \begin{subfigure}{0.18\textwidth}
      \includegraphics[trim={1.5cm 0.8cm 1.0cm 2.1cm},clip,width=\linewidth]{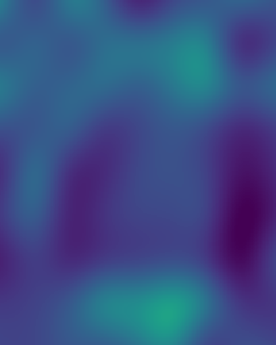}
    \end{subfigure} \hfil 
    \medskip
    
    \begin{subfigure}{0.18\textwidth}
        \includegraphics[width=\linewidth]{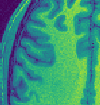}
        \caption{FS\label{fig:sub:FS}}
    \end{subfigure}\hfil 
    \hfil
    \begin{subfigure}{0.18\textwidth}
      \includegraphics[width=\linewidth]{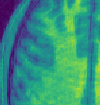}
      \caption{Zero-filled}
    \end{subfigure}\hfil 
    \hfil
    \begin{subfigure}{0.18\textwidth}
      \includegraphics[width=\linewidth]{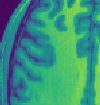}
      \caption{Baseline}
    \end{subfigure} \hfil 
    \hfil
    \begin{subfigure}{0.18\textwidth}
      \includegraphics[width=\linewidth]{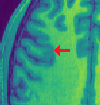}
      \caption{Joint recon.}
    \end{subfigure} \hfil 
    \hfil 
    
\caption{Reconstruction results for R = 3, with (\subref{fig:sub:FS}) the fully sampled image, (b) the zero-filled image, (c) the reconstruction with no bias field estimation, (d) the joint reconstruction with bias field estimation using N4. 
The first three rows show reconstruction results for an HCP image, its zoomed-in version and the corresponding bias field. The next three rows show results for an in-house measured image. 
For visualization purposes, MR images are clipped to [0, 1.2] and bias fields, to [0.5, 1.8].}
\label{visual_recon_HCP}
\end{figure}

To evaluate the reconstruction quality of the different models, we computed, for each image, the percentage RMSE. More specifically, the evaluation metric was given as $\textrm{RMSE} = 100 \times \sqrt{\frac{\sum_i (|\Bb \xb| - |\hat \xb|)^2_i}{\sum_i |\hat \xb|^2_i}}$ where $\hat \xb$ is the original, fully sampled test image, and where the summation was applied pixel-wise. 
The error was computed on the skull-stripped images only in the brain area.

To evaluate the statistical significance of our results, we performed a permutation test~\cite{permutationtest} with 10,000 permutations to assess the null hypothesis that the RMSE's for the reconstruction with and without bias field estimation are from the same distribution. From these tests, we reported the p-values.

\section{Experiments and Results}

To assess the hypothesis that correcting the bias field improves the overall reconstruction quality, we performed reconstructions on the original and modified HCP test sets as well as on the in-house measured images.

When applying our proposed method, we reconstructed undersampled images with a bias field, and used a VAE trained on images without bias field as well as the N4 bias field estimation algorithm. To evaluate our approach, we also ran baseline experiments on the same images, where we used a VAE trained on images with a bias field and did not apply bias field estimation during reconstruction. Given that the ground truth (fully sampled) images naturally have a bias field in them, we utilized the last bias field estimate multiplied with the reconstructed image, i.e. $\Bb\xb$, for visualisations and RMSE calculations.

\begin{table} [H]
    \centering
    \caption{Table of RMSE values. R is the undersampling factor. Numbers indicate the mean (std). The * indicates a p-value of less than 0.05. The baseline method is the DDP algorithm described in~\cite{DDP} that does not explicitly model the bias field. The proposed joint reconstruction method estimates the bias field using N4 and explicitly models it during reconstruction.}
    \resizebox{\textwidth}{!}{\begin{tabular}{ | p{4em} || c | c | c | c || c | c | c | c || c | c | c | c |}
        \hline
        \multirow{2}{4em}{Method} & \multicolumn{4}{|c||}{HCP dataset} & 
        \multicolumn{4}{|c|}{Modified HCP dataset} & 
        \multicolumn{4}{|c|}{In-house measured dataset}\\
    \cline{2-13}
          & R= 2 & R = 3 & R = 4 & R = 5 & R= 2 & R = 3 & R = 4 & R = 5 & R= 2 & R = 3 & R = 4 & R = 5 \\
    \hline
    \hline
        \multirow{2}{4em}{Baseline} & 2.24 & 3.39 & 4.42  & 5.72 & 2.24 & 3.45 & 4.26 & 5.58  & 4.64  & 6.852 & 8.593 & 11.046 \\
         &  (0.31) & (0.45) & (0.51) &  (1.05) & (0.38) & (0.60) & (0.46) & (1.20) &(0.391) & (0.726) & (1.344) & (1.727)\\
    \hline 
        \multirow{2}{4em}{Joint recon.} & 2.27 & 3.34 & 4.35* & 5.52 & 2.20* & 3.33* & 4.16 & 5.12*  & 4.62  & 6.714* & 8.218* & 10.567* \\
         &  (0.34) & (0.40) & (0.47) &  (0.66) & (0.39) & (0.53) & (0.54) & (0.57) & (0.418) &  (0.821) & (1.266) & (1.632) \\
    \hline
    \end{tabular}}
    \label{HCP_rmse_table}
\end{table}

The results in Table~\ref{HCP_rmse_table} indicate that the proposed joint reconstruction method with bias field estimation improves the reconstruction quality in terms of RMSE when the bias field of the test image is different from those in the training set. In these cases, namely the experiments with the modified HCP and in-house measured images, the improvement is statistically significant with a p-value of less than 0.05 for nearly all undersampling factors. For the unmodified HCP dataset, where the bias field in the test images matches those in the training set, we do not expect a big difference in the performance, which is reflected in the results.

The quantitative improvement is also supported by the visual inspection of the images given in Figure~\ref{visual_recon_HCP}. From the HCP image, one can observe that the level of artifacts is reduced with the proposed method. This becomes more evident in the zoomed-in images. The red arrow points to a part of the image where the proposed method can reconstruct the structures faithfully, whereas the baseline method struggles. Aliasing artifacts are globally suppressed better with the joint reconstruction method. Similarly, for the in-house measured image, the grey matter structure that the red arrow points to is not reconstructed in the baseline method, whereas it again appears with the proposed method.

In this work, we demonstrate the performance loss due to the bias field for a specific algorithm, into which we also integrate our proposed solution. However, the problem may not be specific to the algorithm used as it arises from the domain gap, which is a fundamental problem affecting machine learning based methods in general. Furthermore, the proposed method of estimating and explicitly modeling the bias field in reconstruction is also a generic approach, which can be integrated into different algorithms.

\section{Conclusion}
In this paper we proposed and evaluated a method for joint reconstruction and bias field estimation to address variations due to the bias field when reconstructing undersampled MR images. The results indicate that the proposed method improves the baseline method (unsupervised learning based reconstruction algorithm), both in RMSE and visually. The improvements can be attributed to two factors. First, the proposed method allows the VAE prior to learn a simpler distribution, and second, providing the bias field explicitly makes the reconstruction problem easier. In essence, estimating the bias field during reconstruction makes the model less sensitive to differences between the data used to train the model and the test data used during reconstruction.

\end{document}